\documentclass[prl,preprint,showpacs]{revtex4}

\usepackage{amssymb,amsmath,bm}
\usepackage[final]{graphicx}

\newcommand{\modena}[0]{{INFM--$S^3$ and
Dipartimento di Fisica Universit\`a di Modena e Reggio Emilia, 
Via Campi 213/A, 41100 Modena, Italy}}

\begin{document}
\bibliographystyle{apsrev}

\title{Tunneling and Electric-Field Effects on Electron-Hole Localization \\
in Artificial Molecules}

\author{Filippo Troiani}\email{troiani@unimore.it}\affiliation{\modena}

\date{\today}

\begin{abstract}

I theoretically investigate the Stark shift of the exciton goundstate 
in two vertically coupled quantum dots as a function of the interdot 
distance. 
The coupling is shown to enhance the tuneability of the linear optical 
properties, including energy and oscillator strength, as well as the 
exciton polarizability. 
The coupling regime that maximizes these properties results from the 
detailed balance between the effects of the single-particle tunneling, 
of the electric field and of the carrier-carrier interaction.
I discuss the relevance of these results to the possible 
implementation of quantum-information processing based on semiconductor
quantum dots: in particular, I suggest the identification of the
qubits with the exciton levels in coupled- rather than single-dots. 

\end{abstract}

\maketitle


Semiconductor quantum dots (QDs) are quasi-zero-dimensional 
heterostructures, where the strong spatial confinement of 
electrons and holes gives rise to discrete optical 
spectra~\cite{hawrylak:98}.
As compared to natural ones, such ``artificial atoms'' show
a wide tuneability of the structural, electronic and optical 
properties, of the coupling with each other and with external 
fields. 
This makes them an ideal laboratory for the many-body physics, 
where the relative importance of the different kinds of interaction 
can be modified in order to produce regimes which are precluded 
to atomic and molecular physics (see, e.g., 
Refs.~\cite{troiani:02,rontani:01}, and references therein).
In particular, the application of an electric field on single and 
coupled QDs and the resulting quantum confined Stark effect (QCSE)
has been theoretically~\cite{barker:01,sheng:02,janssens:02} 
and experimentally~\cite{fry:00,findeys:01,patane:00,passaseo:01} 
investigated, in order to gain insight into the spatial distribution 
of the carriers and into the effects of possible strain-induced or 
built-in electric fields.

In recent years an increasing attention has been devoted to QDs 
as candidates for a solid-state implementation of quantum-information 
processing (QIP). 
Many possible schemes have been proposed, involving the use of either 
spin~\cite{loss:98} or orbital~\cite{barenco:95} degrees freedom 
as qubits. 
The latter include those that identify the computational degrees of 
freedom with specific exciton levels and propose to manipulate them on 
the ps timescale by means of coherent-carrier control 
techniques~\cite{troiani:00,biolatti:00,chen:01}. 
In each of these QD-based schemes the coupling between dots is a 
crucial issue and is generally supposed to be tuneable by means of
external electric or magnetic fields.
In particular, the use of the electric field has been suggested in 
order to induce relevant dipoles in the exciton states, which would 
result in a dipole-dipole interaction between excitons (qubits) localized 
in neighbouring dots~\cite{biolatti:00}. The more critical aspects of 
similar proposals are related to the need of spectrally 
resolving each single qubit within 
a large array and to the efficient 
decohering phenomena, such as the radiative recombination of the 
electron-hole pairs~\cite{decay}, that result in decoherence times 
between tens and hundreds of ps. 
In these two respects an increased tuneability of the exciton energy 
and of the oscillator strength might represent a relevant improvement: 
one of the main goals of the present work is to demonstrate that this can 
be achieved by switching from single to coupled QDs. In particular, I will 
focus on the 
range of electric-field values that is compatible with finite exciton oscillator strength, as required in order to allow optical manipulation of the exciton states. 


As a first step, I compute the single-particle eigenfunctions and
eigenvalues for electrons and holes within the envelope-function and 
effective-mass approximations~\cite{hawrylak:98}. 
The external confinement of the carriers in the double dot is described 
by a prototypical confinement potential, which is double-box-like 
along $z$ and parabolic in the $(x,y)$-plane; an additional term 
in the single-particle Hamiltonian $H_{SP}$ accounts for the electric 
field $F$, directed along $z$:
\begin{eqnarray}
H_{SP}= \sum_{\alpha = e,h} \int \hat\psi^{\dagger}_{\alpha} ({\bm r})
\Big[-\frac{\hbar^2}{2m^*_{\alpha}} \nabla^2 + V_{\alpha}^{DW}(z) +
\nonumber \\ \frac{1}{2}m^*_{\alpha}\omega^2_{\alpha} (x^2+y^2) -
q_{\alpha}Fz \Big] \hat\psi_{\alpha} ({\bm r}) \, d{\bm r},
\end{eqnarray}
where $\hat \psi_{e}({\bm r})$ and $q_e$ ($\hat \psi_{h} ({\bm r}) $ 
and $q_h$) are the electron (hole) field operators and electric charge
respectively, $ V_{e,h}^{DW}(z) = V_{e,h}^{0} [\theta (|z|-d/2-l)
+ \theta (d/2-|z|)] $, $l=10$~nm and $d$ are the width of the wells and 
the interdot barrier respectively, $V_{e,h}^{0} = 400, 215$~meV are 
the well depths for electrons and holes.  

Starting from these ingredients, I compute the interacting electron-hole
states by means of a full configuration interaction (CI) 
calculation~\cite{banyai:93}. For each number of electron and holes, 
I generally truncate the Hilbert space by keeping only the $N_H$ configurations 
of lowest single-particle energy, $N_H$ beeing determined on the grounds 
of a convergence criterion. Within such subspace, I diagonalize the 
Hamiltonian matrix 
$H_{SP}+H_C$, where 
\begin{equation}
H_C=\frac{1}{2}\sum_{\alpha\beta=e,h} 
\int\int\frac{d{\bm r}\;d{\bm r}'\;} {(\kappa|{\bm r}-{\bm r}'|)}
\hat \psi^{\dagger}_{\alpha} ({\bm r}')
\hat \psi^{\dagger}_{\beta } ({\bm r} ) 
\hat \psi_{\beta}            ({\bm r} ) 
\hat \psi_{\alpha}           ({\bm r}') 
\end{equation}
accounts for the carrier-carrier Coulomb interactions and for 
the resulting correlation effects ($\kappa$ is the semiconductor
dielectric constant, ${\bm r}\neq{\bm r}'$). In the investigation of
the single exciton ground state 20 SP states for both electrons and 
holes are kept and no further truncation of the Hilbert space
is performed (therefore $N_H = 400$ configurations).


As a preliminary step, let me consider the effects of the interplay 
between tunneling and electric field on the SP states, which is  
described by the $z$-dependent part of $H_{SP}$. 
In Fig.~1 I plot the distributions along $z$ corresponding to the
electron and hole SP groundstates as a function of $F$ and of the 
interdot distance $d$.
These eigenstates result from 
a competition between the coherent coupling, that tends to delocalize 
the wavefunctions over the double dot, and the field, that instead 
tends to localize them in one dot or in the other.
While the tunneling-induced energy splitting between the bonding 
and antibonding states is conventionally denoted by $2t$, 
the one that the field induces between states localized in two 
different dots is approximately given by $2\Delta\equiv e(d+l)F$. 
The ratio $x\equiv\Delta /t$ thus provides a good estimate of the 
relative strength of such two effects and of the resulting degree 
of localization of each carrier: this is found to increase 
from electrons to holes (typically $m^*_e < m^*_h$ and thus 
$t_e > t_h$) and with increasing $d$ (see Fig.~1(a-d)). 
In the same range of values of $F$ the SP eigenstates 
and eigenfunctions corresponding to a single dot (not reported 
here) are hardly affected by the field.
Within the present picture, such a difference can be 
understood by thinking of the single QD as a double dot in the 
in the limit $ d \rightarrow 0 $: the values of $ x \equiv \Delta 
/ t $ corresponding to an electric field $ F \leq 8 $~kV/cm turn 
out to be very small both for electrons ($ x_e \leq 0.038$) and 
for holes ($ x_h \leq 0.18 $). 

The structural parameters which determine the carrier confinement in 
each dot, namely $l$ and $\hbar\omega_{e,h}$, are kept constant 
throughout the following calculations: while a detailed understanding
of their role would require a full exploration of the parametric 
space, a few comments can be done on the grounds of the present 
results. In particular, the possible reduction of the well width $l$ 
would mainly result in an enhancement of the carrier penetration in 
the barrier and therefore of the interdot coupling and of $t$, 
whose role and interplay with the electric field are discussed above. 
The in-plane confinement, instead, does not directly affect the 
interdot coupling (and therefore $t$ or $x$). A possible increase 
of $\hbar\omega_{e,h}$ would essentially produce an enhancement of the 
intradot Coulomb matrix elements and therefore of the Coulomb-induced 
effects (see the discussion below).

The properties of the exciton states result from the interplay 
between the abovementioned features of the SP states and the electron-hole 
Coulomb interaction.
In Fig.~2 I plot the Stark shifts and the oscillator strengths 
of the exciton groundstate $ | X_0 \rangle $ as a function of $F$ 
and $d$ (panel (a)) and the contributions to the total energy arising 
from $H_{SP}$ and $H_C$ (Fig.~2(b-c)); 
the results for the single dot are also reported for a comparison. 
Quite generally, the magnitude of the QCSE is greatly enhanced and 
the wavefunctions are much more affected by the field in coupled dots 
as compared to single ones. 
While in single QDs the Stark shift $ \Delta E \equiv E(F)-E(0) $ 
typically shows a quadratic dependence on $F$, 
due to the perturbative nature of the contribution from the field, 
a nonparabolic behaviour is expected in the case of ``artificial 
molecules''~\cite{sheng:02,janssens:02}. 
The field induces a monotonic decrease of the single-particle energy
$E_{SP}$, which determines the negative sign of the Stark shift, 
and a progressive spatial separation of the carriers, as demonstrated
by the increasing Coulomb energy $ E_C $ and oscillator strengths 
(Fig.~2(a-c)).

A closer inspection of the plots reveals the occurence 
of two regimes. In the strong-field region ($ F \gtrsim 6 $~kV/cm) 
the overlap between the electron and hole distributions rapidly 
decraeses and $ \Delta E $ depends quasi linearly on $F$ (as in 
the case of a field-rigid dipole interaction): both trends are 
enhanced at larger values of the interdot distance $d$.
In the weak-field region ($F \lesssim 6 $~kV/cm) the interplay 
between tunneling, Coulomb interaction and electric field is 
more balanced: rather counterintuitively, the energies $E$, $E_{SP}$
and $E_C$ get closer to the single-dot value as $d$ is increased. 
As discussed in the following, this can be attributed to the fact 
that the strength of the interdot coupling determines to which 
extent the carrier-carrier Coulomb interaction can oppose the effects 
of the electric field.

As discussed in Ref.~\cite{sheng:02}, the possible presence of strain 
effects is expected to meaningfully affect the valence-band states and 
thus the optical properties, e.g., in self-assembled InAs/GaAs QDs. 
More specifically, the authors show how the biaxial components of the 
strain field decrease the height of the interdot barrier, increase 
the energy levels of the confined holes, and induce a stronger 
confinement of the holes in the upper dot as compared to the lower one. 
The first two features result in an effective enhancement of the 
interdot coupling for the holes, the latter implies an asymmetry in 
the QCSE with respect to the direction of the applied electric field. 
According to the present analysis, neither of these two effects is 
required for the anomalous Stark shift to occur in coupled QDs:  
I thus expect it to be observed also in case of softened or 
negligible strain fields in and around the dots.

The spatial distribution of the carriers provides some more physical 
insight in the exciton groundstate $ | X_0 \rangle $. 
In Fig.~3(a)[(b)] I plot the probability of finding the electron 
(hole) in the upper dot for $d=$~1, 2 and 3~nm (red, green and blue 
lines), and in the upper part of the single dot (black dotted line). 
As expected from the previous discussion on the SP states, the hole 
is highly localized in the upper dot already at small values of 
the field and especially for large values of $d$. 
On the other hand, the behaviour of the electron again reflects 
the existence of the abovementioned two regimes. 
In the strong-field region the 
electron is seen to approach the lower dot, under the direct 
influence of the field; as observed in Fig.~2 for the energies, the 
derivatives of the curves in this region ($| \partial N_{ud}^{e,h} /
\partial F |$) increase with increasing $d$. 
In the weak-field region the Coulomb interaction with the hole 
dominates on the interaction with the field and the electron tends
to be localized in the upper dot. This spatial arrangement is also
reflected by the small changes in this region of the oscillator 
strength and Coulomb interaction (Fig.~2(b-c)). 
The curves plotted in Fig.~3(a) cross at $ F \simeq 7 $~kV/cm: 
both in the weak- and in the strong-field regimes the degree of 
localization of the electron, in the upper or in the lower dot 
respectively, is opposed by the tunneling. 
For $d=1$~nm the electron is basically frozen on a delocalized bonding 
orbital and $ N^e_{ud} $ does not deviate significantly from $ 1/2 $
in the considered range of $F$ values; at large interdot distances
($d=3$~nm) $t_e$ is small enough for the antibonding orbital to be 
also occupied in order to produce the localization of the electron 
in either dot, depending on the relative strength of the external- 
and of the hole-induced electric field.  

The polarizability of the exciton groundstate reflects this behaviour.
While the value of $D$ is on average greatly enhanced in coupled QDs
as compared to the single one (black dotted vs. coloured lines in 
Fig.~3(c)), the sign of $ \partial D / \partial d $ actually depends 
on the range of electric-field values of interest.
For $ F \lesssim 6 $~kV/cm, where the electron tends to follow 
the hole in the upper dot, the larger degree of localization resulting 
from the larger values of $d$ tends to suppress the dipole;
for $ F \gtrsim 6 $~kV/cm, where the electron interaction with the 
field dominates on that with the hole, a larger degree of localization
results in a larger induced dipole. 
Depending on the detailed balance between SP and Coulomb terms in the 
Hamiltonian, either the strongly or the weakly coupled QDs show the 
highest polarizability.

Let me finally give a rough estimate of the dipole-dipole interaction 
that a field of 8~kV/cm could induce between two pairs of vertically 
coupled dots, given that $ E_{int} \simeq [ \vec{D}_1\cdot\vec{D}_2 - 
3 (\vec{D}_1\cdot\hat{r}_{12}) \, (\vec{D}_2\cdot\hat{r}_{12}) ] / 
\kappa r^3_{12} $, and with $ \vec{D}_1 \parallel \vec{D}_2 \perp 
\vec{r}_{12} $, $ | \vec{D}_{1,2} | =10\; e\times $nm and $r_{12} 
= 20$~nm. 
The energy shift produced by the exciton-exciton interaction is 
 $ E_{int} \simeq 1.4$~meV, i.e. the same order of magnitude of 
the exciton-exciton interactions in a single QD of the same sizes.


In the perspective of the QD-based QIP schemes, the large enhancements 
of the exciton polarizability and of the optical tuneability suggest 
the identification of the computational degrees of freedom with 
exciton levels in vertically-coupled dots rather than in single
ones, as proposed in previous works~\cite{troiani:00,biolatti:00}.
The tuneability of the exciton levels by means of local electric fields 
can be exploited in order to favour the spectral resolution of specific 
transitions (qubits) within an array of neighbouring dots of similar 
dimensions and to bring specific transitions in and out of resonance 
with the laser fields used for the gate implementation.
The polarizability of the exciton states, instead, is required in order 
to induce a relevant dipole-dipole interaction $E_{int}$ between excitons 
that are localized in different and (quantum-mechanically) decoupled dots.
In fact, the value of $ E_{int} $ determines 
a lower bound to the duration $ \tau_g $ of the required laser pulses 
($ \tau_g \gtrsim \hbar / E_{int} $): an exciton-exciton interaction of 
the order of one meV, as the one estimated above, allows in principle
to perform the quantum gates on the ps timescale.

The suggested geometry can be integrated in a planar array of 
vertically coupled dots: each double dot would couple to its nearest 
neighbours only by means of the field-induced dipole-dipole interaction,
while the in-plane SP tunneling should be suppressed in order for the 
required tensor-product structure of the Hilbert space to be 
preserved~\cite{jozsa:97}.
As compared to the ones that consider an in-plane electric 
field, the present geometry has a twofold advantage: 
(i) it avoids the breaking of the optical selection rules related 
to the cylinder symmetry of the dot and the resulting 
thickening of the optica spectra~\cite{banyai:93}, which interferes with the 
identification and selective addressing of the single absorbtion 
lines; (ii) it does not require the etching of a mesa structure
and thus a strong limitation on the number of qubits that compose the 
quantum hardware.

To summarize, I have investigated the effects of a vertical electric 
field on the exciton groundstate and on the related optical properties 
of two vertically coupled QDs. 
The present results suggest the identification of each qubit with the 
exciton groundstate in a pair of coupled dots rather than in a 
single one, in view of the higher degree of tuneability of the linear 
optical properties, including the energy levels and the corresponding 
oscillator strength. 
Besides, the polarizability of the lowest exciton states in 
artificial molecules is enhanced by one order of magnitude or more as 
compared to the one in artificial atoms, resulting in relevant 
dipole-dipole interactions between excitons localized in different 
pairs of coupled dots.
More specifically, due to the interplay between tunneling, 
carrier-carrier Coulomb interactions and electric field, both the 
tuneability of the optical properties and the polarizability are 
maximized either in the strong- or in the weak-coupling regime, 
depending on the range of $F$ values. 

The author is grateful to E. Molinari and U. Hohenester 
for stimulating and encouraging discussion.
This work has been supported in part by INFM through PRA-99-SSQI and
by the EU under the IST programme ``SQID''.


\begin{figure}
\caption{Distribution $ \rho_{e,h} (z) $ of the electron (a,c) 
and hole (b,d) 
eigenstates as a function of the applied field $F$ for $d=1$~nm
(a,b) and $d=3$~nm (c,d), where $\rho_{e,h} (z) = \iint |\psi ({\bm r})|^2 
\; dx\,dy $.
The degree of localization depends on $x\equiv\Delta /t$, 
which is proportional to $F$, ranging from 0 to 8~kV/cm.
The values of the remaining physical parameters
of the physical parameters are the following:
$\hbar\omega_{e,h}=20,3.5$~meV; $l=10$~nm for the well width;  
$V_0^{e,h}=400, 215$~meV for the well depth (band offsets). 
The effective masses and the dielectric constant are the ones 
of GaAs (wells) and AlGaAs (barrier).}
\end{figure}

\begin{figure}
\caption{(a) Quantum confined Stark effect as a function of the 
applied field for vertically 
coupled QDs with interdot distances $d=1,2$ and $3$~nm (red,
green and blue squares respectively) and for the single dot 
(white squares). The sides of the squares are proportional to the 
intensity of the corresponding optical transitions and thus to 
the square moduli of the oscillator strengths. 
The contributions to the exciton energy arising from the single-particle 
($ \langle X_0 | H_{SP} | X_0 \rangle $) and from the Coulomb terms 
($ \langle X_0 | H_{C} | X_0 \rangle $) in the Hamiltonian are plotted 
in the (b) and (c) panels respectively, with the same convention for 
the colours.}
\end{figure}

\begin{figure}
\caption{Probability of finding the electron (a) and the hole (b)
in the upper dot for $d$= 1, 2 and 3~nm (red, green and blue lines 
respectively), where $ N^{e,h}_{ud} = \langle X_0 | \int \;\hat 
\psi^{\dagger} ({\bm r}) \;\hat\psi ({\bm r}) \; d{\bm r} | X_0 \rangle $;
the black dotted line is the probability of finding the carriers 
in the upper half of the single dot.
(c) Dependence on $F$ of the induced dipole $ D = \langle X_0 | 
\int z \; ( \; \hat{n}_h - \hat{n}_e ) \; d{\bm r} \; | X_0 \rangle $, 
with the same convention for the colours. }
\end{figure}

\setcounter{figure}{0}
\newpage
\begin{figure}
\centerline{\includegraphics[width=0.85\columnwidth,height=16cm]{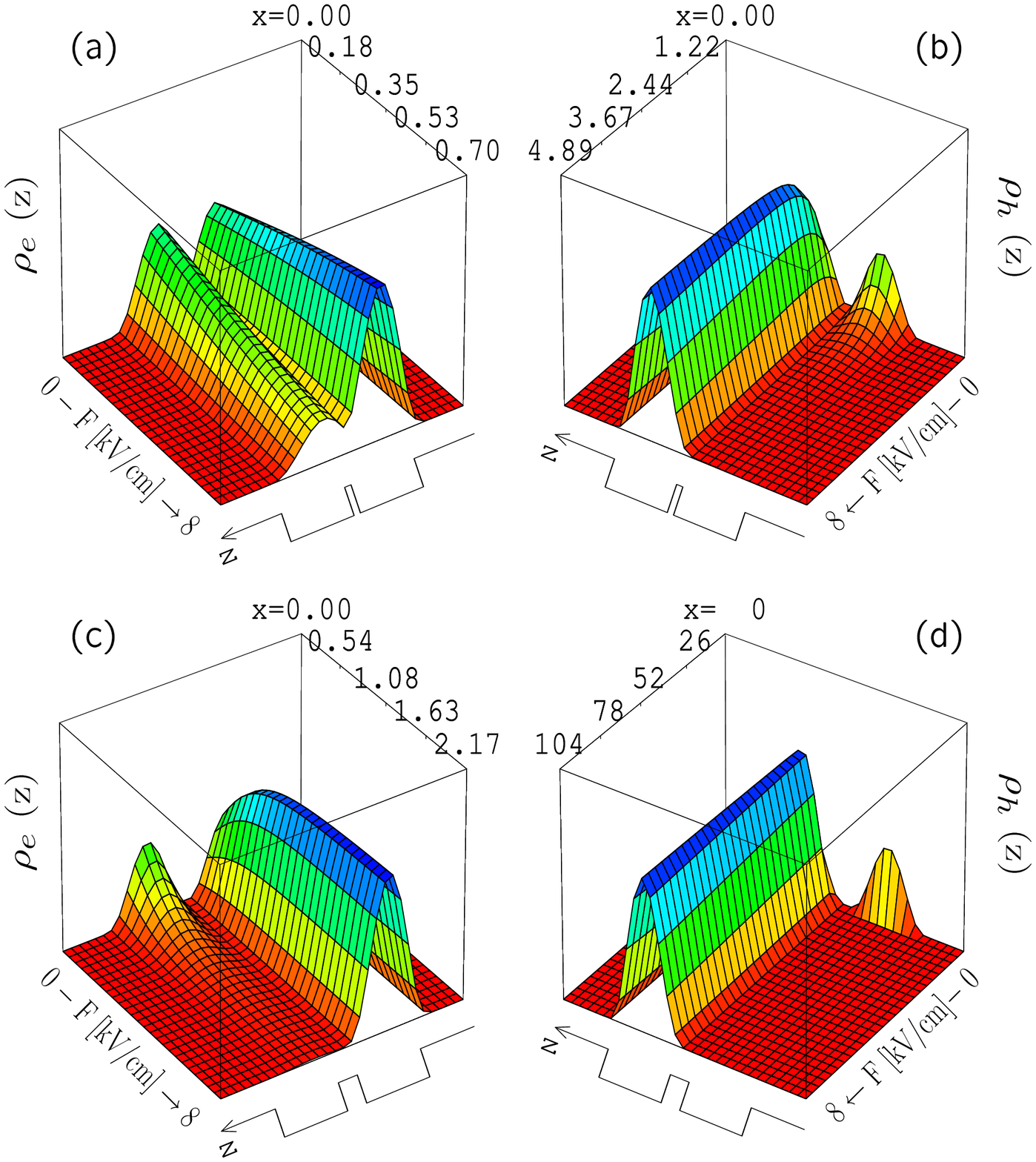}}
\caption{F. Troiani, Tunneling and Electric-Field Effects on Electron-Hole Localization in Artificial Molecules}
\end{figure}

\newpage
\begin{figure}
\centerline{\includegraphics[width=0.85\columnwidth,height=16cm]{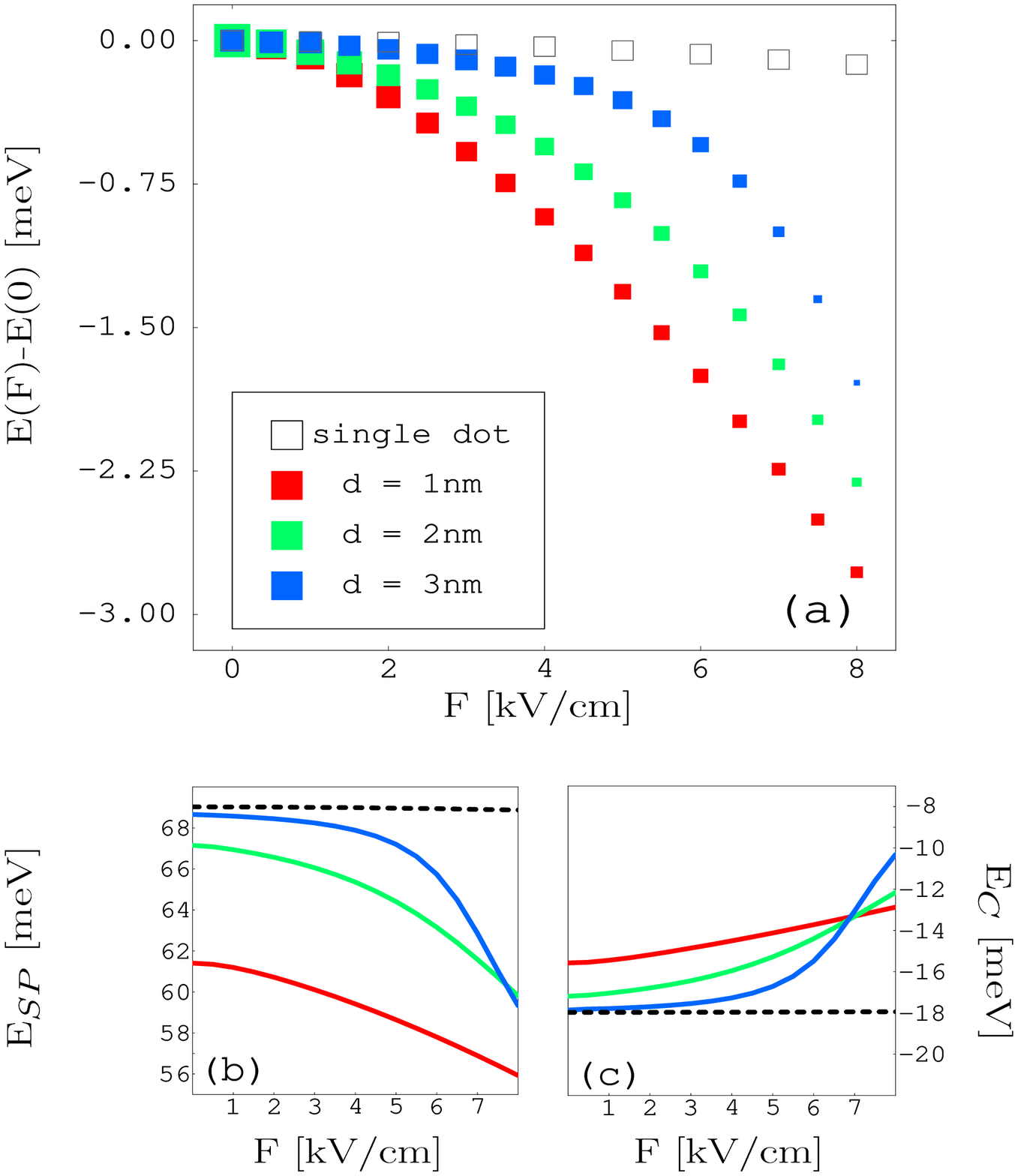}}
\caption{F. Troiani, Tunneling and Electric-Field Effects on Electron-Hole Localization in Artificial Molecules}
\end{figure}

\newpage
\begin{figure}
\centerline{\includegraphics[width=0.85\columnwidth,height=20cm]{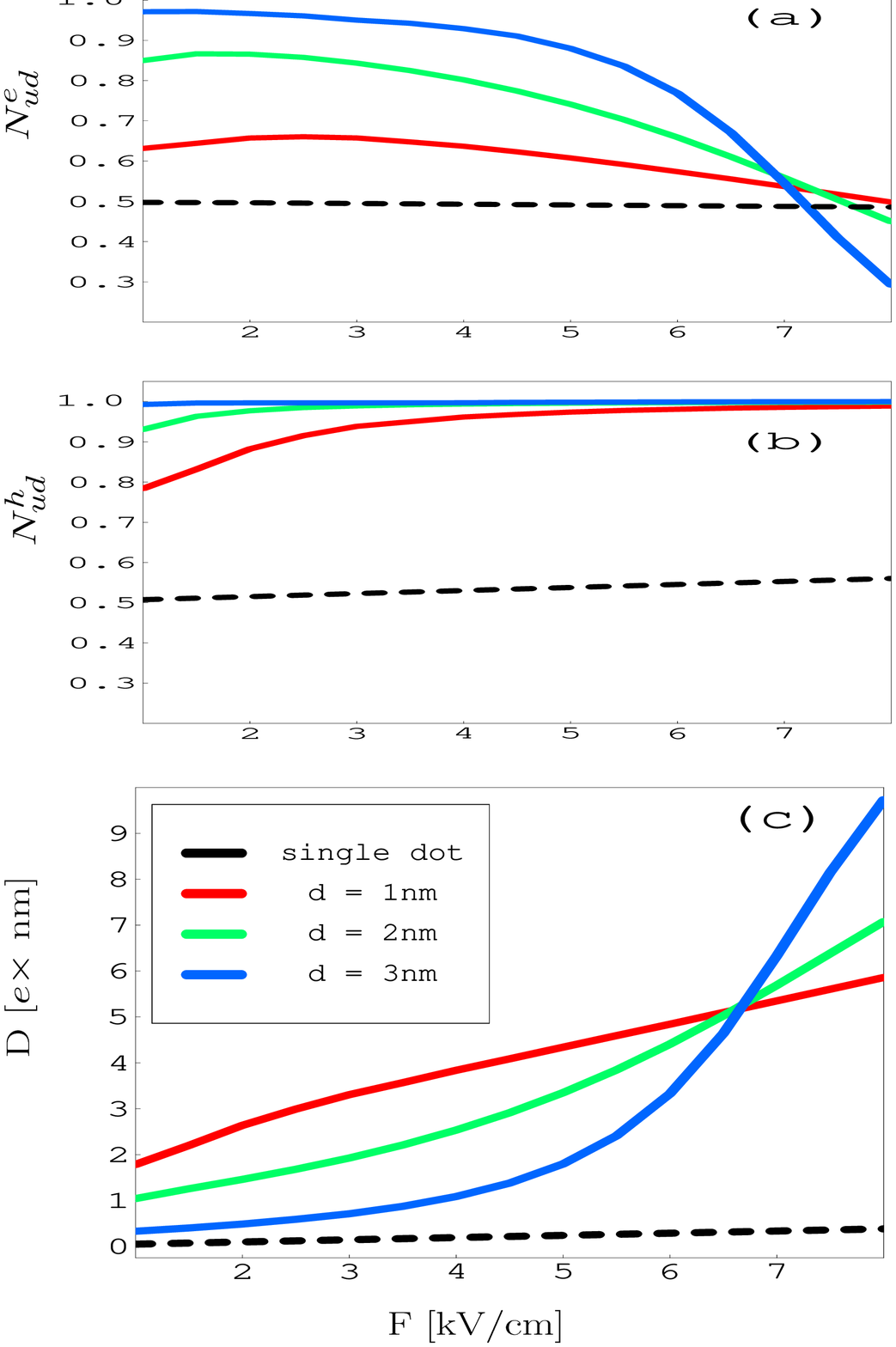}}
\caption{F. Troiani, Tunneling and Electric-Field Effects on Electron-Hole Localization in Artificial Molecules}
\end{figure}

\end{document}